# The Solar Orientation of the Lion Rock Complex in Sri Lanka

**Amelia Carolina Sparavigna**
**Department of Applied Science and Technology, Politecnico di Torino, Italy**

**Abstract** This paper discusses the solar orientation of the archaeological complex of Sigiriya, the Lion Rock, in Sri Lanka. We can see that the axis of this complex is oriented with the sunset of the zenithal sun.

1. **Introduction**

Several ancient ceremonial structures are designed to align with the repeating patterns of sun or moon or even stars. We have the Stonehenge megalithic monument for instance, which has alignments with summer and winter solstices, and the Karnak temple in Egypt, aligned again with the sun on solstices [1,2]. Even the gothic cathedrals have a solar orientation [3]. These huge buildings have their axis aligned with the azimuth of the sunrise on a given day of the year, probably the day of their foundation.

At latitudes above the tropical zone, the sun reaches the highest noon altitude on the summer solstice and the lowest one on the winter solstice. In any case, this angle is below 90 degrees. But, when we are in the tropical zone the sun reaches the zenith, that is an altitude of 90 degrees. In this paper, we will show that, besides the alignment with the sunrise and sunset azimuths on solstices, we can have in the tropical zone, an alignment with the zenithal sun, that is a design of the site with the sunrise or sunset azimuths of the day during which the sun reaches the zenith. An example of this alignment is the Lion Rock complex in Sri Lanka.

2. **The Lion Rock complex**

The ruins of a huge palace built by King Kassapa I (477–495 CE) are on the top of a granite rock, known as Sigiriya, the Lion Rock [4-8]. This site is in the heart of Sri Lanka, dominating the neighboring plateau, inhabited since the 3rd century BC, and hosting some shelters for Buddhist monks [4]. A series of galleries and staircases, having their origin from the mouth of a gigantic lion made of bricks and plaster, provide access to the ruins on the rock. In the Figure 1, it is possible to see the site surrounded by a wall and the rock inside.

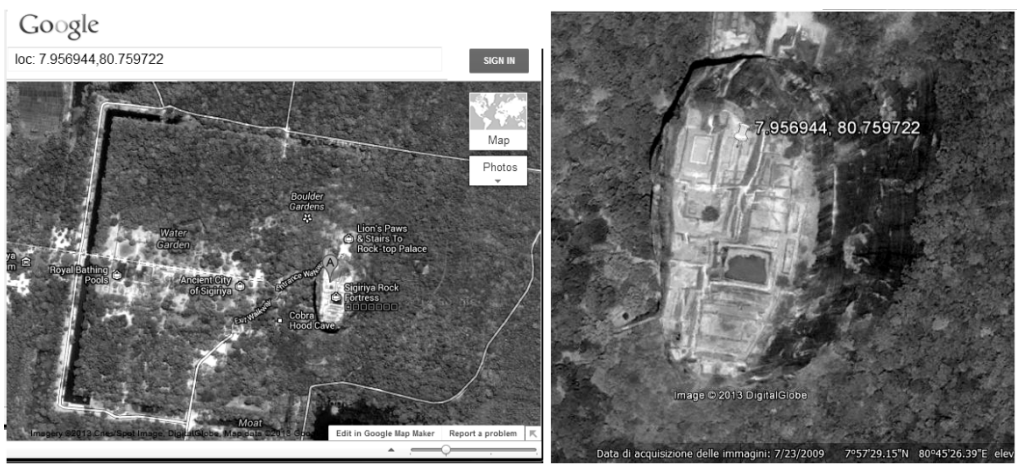

**Figure 1 – The Sigiriya complex as we can see in the satellite maps. On the right the Lion Rock.**

Sigiriya is a unique witness to the civilization of Sri Lanka during the years of the reign of Kassapa [5]. The site is rich of frescoes, which originated a pictorial style used for many centuries. However, the fame of the site is mainly due to the fact that Kassapa I established his capital there in a fortified palace. After the death of Kassapa, the site of Sigiriya returned to the monks, and then was progressively abandoned.

At the summit of the rock, there is the fortified palace with its ruined buildings, cisterns and rock sculptures. At the foot of the rock we find the lower city surrounded by walls. The eastern part of it has not yet been totally excavated. The western aristocratic part of the capital of Kassapa I was embellished by terraced gardens, canals and fountains.

The Gardens of the Sigiriya city are an important characteristic of the site. They are divided into three distinct forms: the water gardens, the cave and boulder gardens, and the terraced gardens [4]. The water gardens are in the central section of the western precinct. They were built according to an ancient garden form, of which they are the oldest surviving examples.

The water gardens are connected with the outer moat on the west and the large artificial lake to the south of the Sigiriya rock. All the pools are also interlinked by an underground conduit network fed by the lake, and connected to the moats.

### 3. Solar orientation

In the Ref.4 it is told that the water gardens are built symmetrically on an east-west axis. In fact, the design of the gardens is symmetrical, however the axis is not oriented on the cardinal east-west line: the site is inclined of 9 degrees, as we can measure from satellite maps (Figure 2).

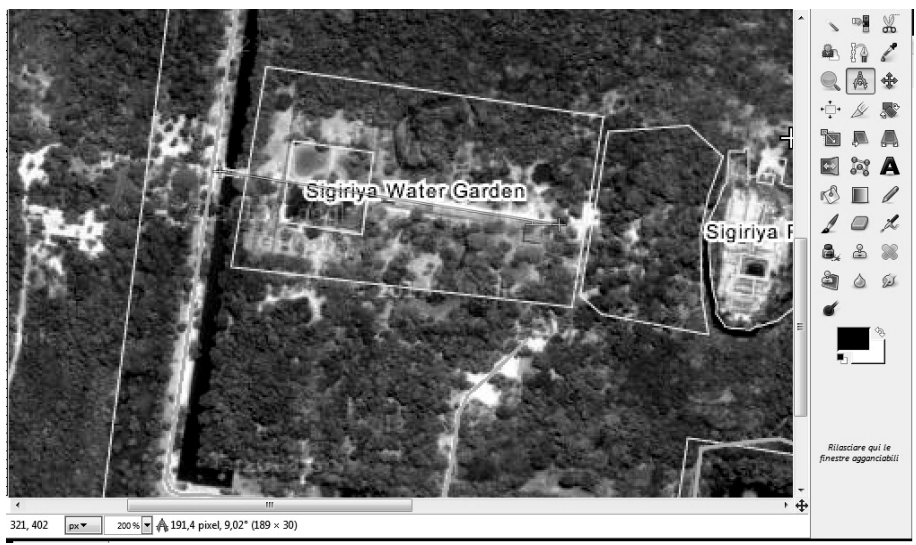

Figure 2 – Measurement of the angle using the GIMP compass.

Since this angle is not negligible, it can correspond to a specific azimuth of the sun.

Let us remember that the azimuths are formed by the vector from the observer to the sun rising or setting on the horizontal plane and a reference vector on this plane. There are several web sites that allow knowing the azimuth and the noon altitude of the sun and moon at a specific location on a given day of the year. For instance, one is the site in Ref.9. Using it, we can obtain at Sigiriya, the following data for the noon altitude and sunset azimuths given in the following table. We see that we have the zenithal sun on April 9 and on the First of September.

| Date | Noon Altitude | Sunset Azimuth |
|---|---|---|
| April 8 | 89.4° | 277.6° |
| April 9 | 89.6° | 278.0° |
| April 10 | 89.5° | 278.4° |
| August 30 | 88.9° | 278.9° |
| September 1 | 89.2° | 278.2° |
| September 2 | 89.1° | 278.0° |

We have that the azimuth is of 8 degrees with respect the cardinal east-west direction. There is then the difference of one degree with the measured angle of the axis of the gardens.

We can also obtain the data on azimuth and noon altitude form a web site that we have already used in some papers (see for instance Ref.10 and references therein): it is that of Sollumis.com [11]. This site allows drawing on the Google satellite maps some lines which show the direction and height of the sun throughout the day. Thicker and shorter lines mean the sun is higher in the sky. Longer and thinner lines mean the sun is closer to the horizon. Using Sollumis.com for instance, we can easily find solar orientations in the layout of some Chinese Pyramids burial complexes [10].

Let us use Sollumis.com on the site of the Lion Rock. The results obtained are shown in the Figure 3. Here we find a sunset azimuthal angle of 9 degrees with respect the cardinal east-west axis, in agreement with the measured angle (see the Figure 2).

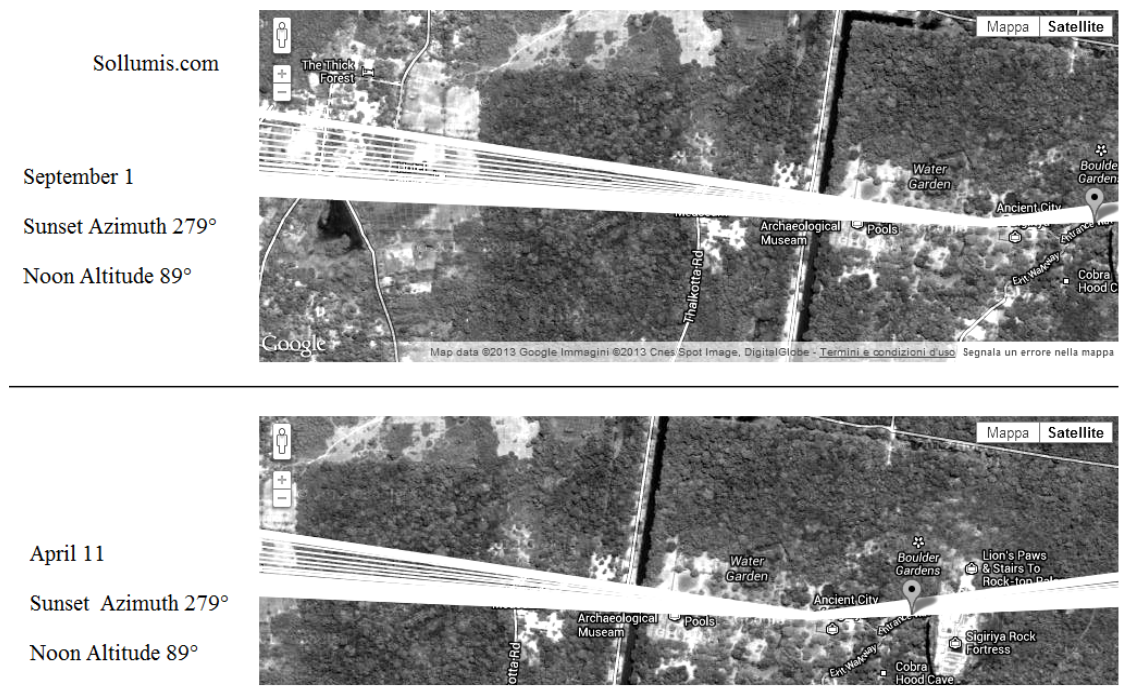

**Figure 3 – The direction of the sun during September 1 and April 11, given by Sollumis.com at Sigiriya. This site provides a polar diagram, overlaying a satellite map, showing the directions of the sun for any day of the year. The lines on the drawing show the direction and altitude of the sun. In the image it is given the sunset azimuth of 279 degrees, which corresponds to an angle of 9 degrees with respect the cardinal east-west direction. The highest value of the noon altitude, all over the year, given by the software is 89°. We can then suppose a truncation of the true value.**

## 4. Conclusion

After this analysis on satellite images and azimuths, we can conclude that the Sigiriya complex was planned with respect of an axis oriented with the sunset of the zenithal sun, that is, oriented with the sunset of a day when the sun reaches the zenith. This fact seems to indicate that, besides the solstices, in the tropical zone the zenithal sun had a ritual importance too.